\documentclass[runningheads]{llncs}
\usepackage{graphicx}
\usepackage[hidelinks]{hyperref}
\hypersetup{
 pdftitle={Intrinsically Motivated Autonomy in Human-Robot Interaction: Human Perception of Predictive Information in Robots},
 pdfsubject={In this paper we present a fully autonomous and intrinsically motivated robot usable for HRI experiments.
We argue that an intrinsically motivated approach based on the Predictive Information formalism, like the one presented here, could provide us with a pathway towards autonomous robot behaviour generation, that is capable of producing behaviour interesting enough for sustaining the interaction with humans and without the need for a human operator in the loop.
We present a possible reactive baseline behaviour for comparison for future research. Participants perceive the baseline and the adaptive, intrinsically motivated behaviour differently.
In our exploratory study we see evidence that participants perceive an intrinsically motivated robot as less intelligent than the reactive baseline behaviour. We argue that is mostly due to the high adaptation rate chosen and the design of the environment. However, we also see that the adaptive robot is perceived as more warm, a factor which carries more weight in interpersonal interaction than competence.},
 pdfauthor={Marcus M. Scheunemann, Christoph Salge, Kerstin Dautenhahn}, 
 pdfkeywords={Robotics, Sustained Interaction, Human-Robot Interaction, Autonomous Human-Robot Interaction, Robot Behaviour, Cognitive Robotics, Robot Control, Autonomous Robots, Intrinsic Motivation, Predictive Information, Information Theory.} 
}
\usepackage{xcolor}

\usepackage{nameref}
\makeatletter
\let\orgdescriptionlabel\descriptionlabel
\renewcommand*{\descriptionlabel}[1]{%
  \let\orglabel\label
  \let\label\@gobble
  \phantomsection
  \edef\@currentlabel{#1}%
  \let\label\orglabel
  \orgdescriptionlabel{#1}%
}
\makeatother

\newcommand{\HypoDistinguishable}{The reactive and the adaptive robot's strategy is described the same.}
\newcommand{\HypoMedianChangeAni}{The median change for the GodSpeed's factor Animacy is zero.}
\newcommand{\HypoMedianChangePInt}{The median change for the GodSpeed's factor Perceived Intelligence is zero.}

\newcommand{\HypoMedianChangeComp}{The median change for the RoSAS' factor Competence is zero.}

\usepackage{csvsimple}

\usepackage{booktabs}

\usepackage[product-units=single, range-units=single, list-units=single]{siunitx}

\begin{document}
\title{Intrinsically Motivated Autonomy in Human-Robot Interaction: Human Perception of Predictive Information in Robots}
\titlerunning{Human Perception of Predictive Information in Robots}
%
\author{Marcus M. Scheunemann\inst{1} \and
Christoph Salge\inst{1} \and
Kerstin~Dautenhahn\inst{1,2}}
\authorrunning{MM. Scheunemann et al.}
%
\institute{University of Hertfordshire, 
AL10 9AB, United Kingdom
\and
University of Waterloo, N2L 3G1, Canada\\
\url{https://mms.ai/sHRI} \\
\email{marcus@mms.ai}
}
\maketitle              
\begin{abstract}
  In this paper we present a fully autonomous and intrinsically motivated robot usable for HRI experiments.
  We argue that an intrinsically motivated approach based on the Predictive Information formalism, like the one presented here, could provide us with a pathway towards autonomous robot behaviour generation, that is capable of producing behaviour interesting enough for sustaining the interaction with humans and without the need for a human operator in the loop.
  We present a possible reactive baseline behaviour for comparison for future research.
  Participants perceive the baseline and the adaptive, intrinsically motivated behaviour differently.
  In our exploratory study we see evidence that participants perceive an intrinsically motivated robot as less intelligent than the reactive baseline behaviour.
  We argue that is mostly due to the high adaptation rate chosen and the design of the environment.
  However, we also see that the adaptive robot is perceived as more warm, a factor which carries more weight in interpersonal interaction than competence.

\end{abstract}
\section{Introduction}
\label{sec:introduction}

Why use autonomous robots for human-robot interaction~(HRI) experimentation~\cite{goodrich2008human} instead of teleoperation by human experimenters or scripted behaviour? Scripting reduces the adaptability of the robot to novel situations, limiting the range and flexibility of interaction scenarios. Teleoperation offers more flexibility, but has problems with scalability, introduction of human bias, and experimenters struggling with acting from the robot's perspective~\cite{adamides2015usability}. In contrast, autonomous robots could realise experiments where they freely interact with humans in real environments. Here we want to introduce some exploratory experiments with an intrinsically motivated robot as a pathway towards realising autonomous robots that are interesting to interact with. 

This paper is motivated by the authors' work with a spherical robot~\cite{ScheunemannDautenhahnEtAl-16}~(similar to Fig.~\ref{fig:robot_environment}) 
and children. The children's interaction patterns were usually very diverse, making a general, pre-scripted robot behaviour for a group of children hard to achieve. Without human interaction or without remotely controlling the robot, the behaviour of the robot was limited, leading to the children losing interest in the robot. While it would be trivial to have a self-controlled robot that exhibits some form of behaviour, the hard question is: What kind of behaviour makes a robot interesting to interact with? We assume that robots are more interesting to interact with if they have perceived agency, allowing the human interaction partner to assign motivations to the robot, support or hinder its goals, or even sympathise with its ``joy'' when achieving a goal. 
Once we identify something as an agent, we are likely to direct our attention towards that agent, trying to understand its goals, intentions and behaviour. 
This interest is, if not synonymous with engagement, a step towards more engagement with the agent.  
While the simple act of moving~\cite{Dautenhahn-97} or the look of a robot's ``head''~\cite{4107851} can change a human's perception of the robots, e.g., animacy, the behaviour also needs an observable goal-directedness~\cite{FukudaUeda-10}.
We think that behaviour based on actual intrinsic motivation~\cite{OudeyerKaplan-09} would be a good candidate to create a consistent perception of a robot's motivation. 
We think that this enables the possibility of eventually sustaining the interaction with a fully autonomous robot.

In psychology, intrinsic motivation is defined as doing an activity for its inherent satisfaction rather than for some separable consequence or reward~\cite{deci1985intrinsic,RYAN200054}. The concept has been linked to the idea of autonomy and agency~\cite{ryan1993agency}, and intrinsic motivations are alternatively defined as those motivations that are an integral, non-instrumental, non-optional part of an agent~\cite{oudeyer2008intrinsic}.
The recent interest in computational approaches to intrinsic motivations~\cite{OudeyerKaplan-09} gives us a rich assortment of formalisms to consider. Most of them have a set of common properties, such as semantic independence, universality and sensitivity in regards to embodiment, and a high degree of robustness. 
They are usually used to answer questions, such as what is a good general heuristic (i.e. motivation function) if my robot knows nothing about the world or even its own morphology? Ideas such as curiosity or self-maintenance are turned into AI formalism that enhance model learning and were generally used to enhance AI and robot performance~\cite{barto2013intrinsic}. More relevant to our approach, recent work in the HCI domain of games used intrinsically motivated agents to generate more interesting, self-learning agents~\cite{Merrick2009}, or to create believable, generic antagonists~\cite{Guckelsberger2018} and companions~\cite{Guckelsberger2016c}.

\section{Predictive Information}
\label{sec:PI}
The Intrinsic Motivation used to generate the robot's behaviour in our study is the Predictive Information~(PI) formalism, closely following the implementation of Martius et al.~\cite{MartiusDerEtAl-13}. A formal introduction of the measures is omitted due to space constraints. Conceptually, the measure falls into a family of learning rules related to the reduction of the time prediction error in the perception-action loop of the robot. The book Playful Machines~\cite{DerMartius-12} offers a good introduction. The book also shows how these approaches can be computed from a robot's perspective alone, and the zoo of different robots and their behaviours presented within shows how behaviour resulting from the different formalisms is sensitive towards the agent's specific embodiment.

Predictive Information~\cite{Ay2008b} derives a specific learning rule, that aims to maximise the mutual information between a robot's past and future sensor states. The relevant literature argues that this produces exploratory behaviour sensitive to the robot's embodiment. The derivation from information theory also offers an intuitive interpretation of the robot's adaptation towards being able to reliably predict its own future from its past, while enriching the diversity of its experiences. The approach we use here \cite{MartiusDerEtAl-13} works by updating the internal neural networks of the robot, one that generates behaviour from sensor input, and the other predicting the futures states. The continuous adaptation, aimed at improving the time-local Predictive Information, moves the robot through a range of behavioural regimes. Importantly, the changes in behaviour are partially triggered by the interaction with the environment, as mediated through the robot's embodiment.
The rate at which those internal neural networks are updated is the one model parameter which we will change between experiments.

To the best of our knowledge, this is the first experiment that uses Predictive Information in the context of HRI, and evaluates how the behaviour based on this intrinsic motivation is perceived by participants. 

\section{Study Design and Procedure}
\label{sec:study}
The challenge in designing this study is that this work is, as far as we know, the first HRI experiment of a robot using Predictive Information. Consequently, we lack an existing baseline for comparison.

\subsection{Baseline Behaviour}
\label{sec:baseline}
We considered the following four alternative means of behaviour-generation for serving as a baseline: (1)~human remotely controls the robot, (2)~random behaviour and (3)~pre-adapted reactive behaviour.

Ideally, we want to see how the algorithm compares to a human remotely controlling the robot. However, human controlled behaviour has a high degree of variance, dependent on the particular human controller. Furthermore, it is unclear how much access the human controller should have to environmental information.
If the human can directly observe the participants, it could obtain much more information than the robot, giving it an unfair advantage in creating behaviour responsive to the participant. 
If we limit the human controller to only the robots' sensors, then the human controller would likely struggle to make sense of this limited input, potentially being unable to control the robot at all. 

The problem with using random behaviour as a baseline is that ``randomness'' actually has a set of parameters that need to be chosen, like how often do values change, or is it the change of value or the value that is being randomised. PI does not show a real pattern of behaviour switches we could use for the timing of changing speed, heading and/or the overall behaviour. We performed some preliminary trials with random values, but were quickly facing the question of a fair baseline behaviour again. Having the experimenter choose these values leads to basically designing a certain kind of behaviour (chosen from a whole range of behaviours), which makes it problematic as a baseline.
We decided to use a pre-adapted reactive behaviour. The pre-adaptation is done with the very same PI implementation and parameters, using the same sensors as the robot will use during the experiments.

\subsection{Robot, Environment and Tasks}
\begin{figure}
\includegraphics[width=0.31\columnwidth]{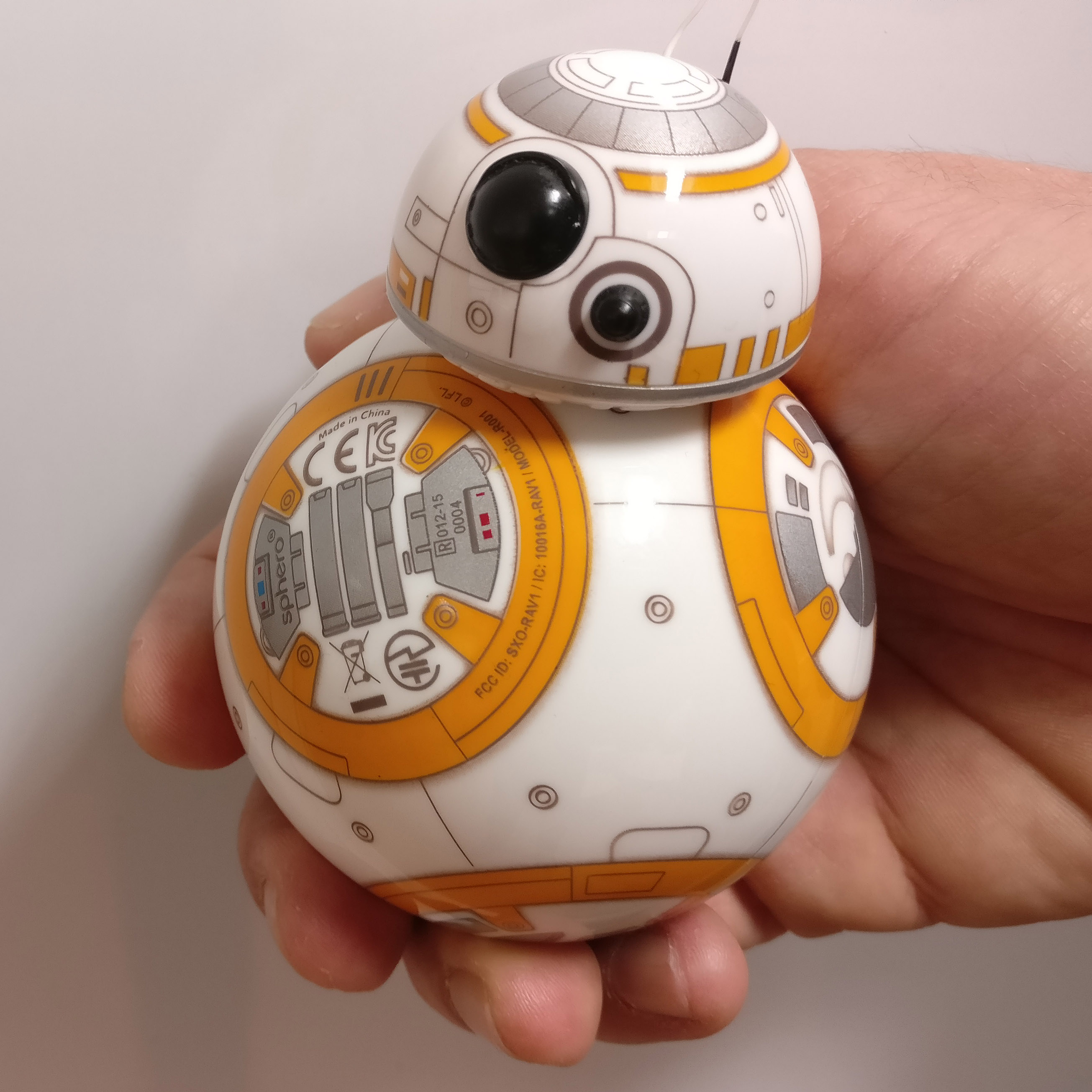}
\includegraphics{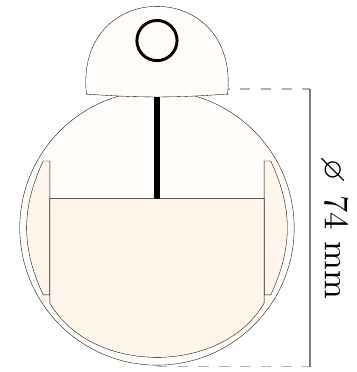}
\includegraphics[width=0.38\columnwidth]{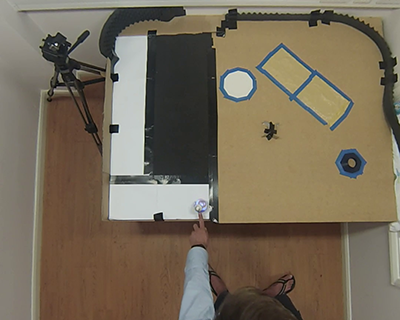}
\caption{Left: The used robot platform BB8 from Sphero. Middle: A 2-D cross-sectional view of the robot. A two-wheel vehicle, kept in position by a heavy weight, moves the sphere when driving. A magnet attached to the vehicle keeps the head on top of the sphere facing in moving direction. Right: The environment the robot explores during the trials from a birds eye perspective. The white area is paper, the black is foam material and the beige coloured area is wood.
At the top of the foam material is a hill area and a pit in the lower part. The bottom edge does not have a wall, forcing the participant to interact with the robot.}  
\label{fig:robot_environment}
\end{figure}
\noindent We want a very simplistic platform with a few degrees of freedom for focusing on the effects induced by Predictive Information.
However, for rich behaviour, the robot also has to offer some sensor capabilities. 
We chose the off-the-shelf spherical robot from the company Sphero, specifically, the BB8 platform as depicted in Figure~\ref{fig:robot_environment}~(left). We chose this version of Sphero because of the advantages the head offered. A magnet keeps the head in driving direction, which gives the user a sense of the robot's direction. This, and the fact that many people know the robot from movies, provides a better impression of a robot than using a white Sphero.
The robot weighs \SI{168}{\gram}. It has a \SI{75}{\mega\hertz} ARM Cortex M4 on board powered by two \SI{350}{\milli\ampere\hour} LiPo batteries.
The robot has a two-wheel electric vehicle inside the spherical shell, as depicted in the cross-sectional view of the robot in the middle of Figure~\ref{fig:robot_environment}.
This is kept in position by a heavy weight, which is made out of the coil for power inductive charging. A connection of a magnet to the vehicle keeps the attached head in position.

There are two ways of controlling the robot. The vehicle inside the robot has two servos. You can either control the speed and direction of each servo directly or you can use Sphero in ``balancing mode''. In this mode the in-built controller tries keeping the robot upright and listens to speed and heading commands.
The direct control mode offers a wide range of possible behaviours, e.g. turning on the spot or more wobbly locomotion. We decided however to use the balanced mode, again, for the sake of simplicity. 

The BB8 can stream sensor information. It offers readings of a 3-DOF~accelerometer, a 3-DOF~gyroscope and the current servo position and servo speed.
It also offers IMU readings in quaternions or euclidean angles\footnote{We found that the roll angle readings of the IMU are faulty.}.
The JavaScript API for BB8 has been unsupported since 2016-05-11 and we decided for developing a custom API which is based on C++ and is available from~\cite{Scheunemann-19suppl}.
Figure~\ref{fig:robot_environment}~(right) shows the experimental environment. Two tables form the space where the robot can move around. The area is \SI{180x120}{\centi\metre} in size. It is open to one side where the participant is supposed to stand and interact with the robot. In Figure~\ref{fig:robot_environment} you can see a participant nudging the robot.
The surface of the table differs in friction and height. The black foam area has a hill (top) and a pit (bottom). Additionally, the black area and the white paper area is softer and has higher friction compared to the wooden part. 

The participants' task is to observe the robot and understand whether it has a strategy for exploring the environment. 
We wanted to encourage the participants to interact with the robot. Therefore, we kept one side open so participants had to interact actively with the robot to prevent it from falling off the table.
We hoped this enforcement of interaction would provide the participants with a better understanding of the robot's capabilities and behavioural richness.
The robot itself has no pre-coded task.

\subsection{Groups and Conditions}
\label{sec:condition}
We decided for two different conditions:
\begin{description}
\item[$REA$ (reactive):] participants interact for approximately \SI{10}{\minute} with a reactive robot and are asked about what they have seen.
\item[$ADA$ (adaptive):] same as $REA$, but the robot is continuously adapting, based on maximisation of Predictive Information as a motivation to interact with its environment.
\end{description}
The adaptive robot in the $ADA$ condition realises behaviour motivated by maximising predictive information, and it continuously updates its internal networks based on that gradient during the experiment. The reactive robot in the $REA$ condition starts with the same networks as the adaptive one (based on pre-trial adaptation), which determines how it reacts to sensor input, but it does not further update its internal network during the experiment.
 
We assign participants into two groups: (A)~$ADA \rightarrow REA$ and (B)~$REA \rightarrow ADA$. The order of $REA$ and $ADA$ is randomly assigned, but balanced over the number of participants.
The starting configuration for both conditions ($REA$ and $ADA$) was generated in two steps. Firstly, we conducted three trials with the robot for \SI{5}{\minute} in the previously described environment. At the end of each trial, we saved the robot's network configuration. In a second step we randomly choose one of these network configurations as the starting configuration. 

The PI formalism allows for having different levels of adaptivity to changing environments and new stimuli.
The update rate for $ADA$ was determined empirically. We noticed that the robot can get caught in the pit we mentioned earlier.
If the robot gets caught in the pit, it would need to adapt to leave and continue exploration. The $ADA$ adaptation rate was set so that the robot would change its behaviour and leave the pit in less than \SI{20}{\second}. 
As we will discuss later, we hypothesised that a high adaptation rate yields a higher perceived intelligence, as the robot would continuously adapt to new stimuli and change the way it would react to certain inputs.

\subsection{Robot's behaviour}
\label{sec:condition_behaviour}

The chosen sensors will determine the behaviour to a large extent, as Predictive Information tries to excite sensor input. For example, if you decide to only use the IMU reading of the yaw angle speed, the robot only needs to adapt its heading in order to excite the sensor. It won't generate any rolling movement. Empirically, we decided for the pitch and roll angles, the $x$ and $y$~component of the accelerometer, as well as the $z$~component of the gyrometer as sensor input.

The aforementioned strategy for using a fixed network for the reactive robot yields a somewhat predictable behaviour for the condition $REA$. The robot prefers left turns in light of environmental perturbations or human interaction, i.e. if it hits a wall, it will almost always turn left. Its major trajectory is that of circling in different radii. With this in mind, we assumed that the same arguments as discussed in section~\ref{sec:baseline} may hold true and people get bored very quickly. However, almost all participants did not recognise the mentioned pattern. Videos for both conditions are available from~\cite{Scheunemann-19suppl}.

The adaptive robot starts with the same network configuration. 
We chose a very high update rate for its model, as discussed in section~\ref{sec:baseline}. 
Its trajectory has a tendency of being straight, if it reaches an obstacle it adapts its heading to be able to continue moving in another direction. 
However, as soon as a participant interacts with the robot, it is not trivial to understand what the robot will do next to increase sensory stimuli.

\subsection{Goals and Hypotheses}
\label{sec:hypotheses}
This paper aims to answer the research questions: (1) Is an adaptive robot perceived as more competent/intelligent and animal-like than the reactive robot? and (2) do the reactive and the adaptive robot have distinguishable behaviours?
We formulate the following null hypotheses for further investigation:
\begin{description}
  \item[$H_0(1)$\label{itm:H_distinguishable}]{\HypoDistinguishable}
  \item[$H_0(2)$\label{itm:H_change_animacy}]{\HypoMedianChangeAni}  
  \item[$H_0(3)$\label{itm:H_change_comp}]{\HypoMedianChangeComp}
  \item[$H_0(4)$\label{itm:H_change_PInt}]{\HypoMedianChangePInt}
\end{description}
\subsection{Measures}
\label{sec:measures}
We decide to use two standardised questionnaires to compare results with other studies: the GodSpeed scale~\cite{BartneckKulicEtAl-09}, which been widely used in many experiments, and the Robotic Social Attributes Scale~(RoSAS)~\cite{CarpinellaWymanEtAl-17}, which is relatively new and has seen little use in HRI so far.

GodSpeed uses a 5-point semantic differential scale and investigates for the factors Anthropomorphism, Animacy, Likeability, Perceived Intelligence and Perceived Safety. 
The authors of RoSAS do not recommend a specific scale, but recommend having a neutral value, e.g. uneven number of Likert elements. We decided to use a 7-Likert scale. It tests for the factors Warmth, Competence and Discomfort. Although we are mostly interested in the factors Animacy, Perceived Intelligence, Competence and Warmth, 
we use all the provided items of both scales. This is done to hide the questionnaire intention and to check for other effects in later research. We will show results for the additional factors Antropomorphism, Likeability, Perceived Safety and Discomfort for completeness purposes, but we will not discuss them in detail in this paper due to space constraints.

We use these scales for the questionnaires after each condition and we ask two open ended questions in addition: (1) ``Can you describe the different behaviours of the robot? Did the robot have any particular strategy for exploring?'' and (2) ``What were the best and/or worst aspects of the robot’s behaviour?''.

\subsection{Methodology}
\label{sec:methods}
Participants are welcomed to the experimental room, they are handed an information sheet and are asked to sign an informed consent form. Then the environment and the robot is presented and briefly described. Participants are informed that the robot's aim is to explore the environment.
They are asked to observe whether the robot follows a particular strategy to do so, and if they can identify any specific behaviour. 
They are also asked to prevent the robot from rolling over the open edge. 
They are shown how to use the hand as a ``wall'' or nudge the robot to prevent it from falling off the side of the arena that is not enclosed by a wall, or to illicit new behaviour through interaction.
Participants then fill in the pre-questionnaire. This gathers information regarding their sex, age and background. Next, the two conditions are presented to the participants on a randomised order. Each lasts approximately \SI{10}{\minute}. They fill in two post-questionnaires containing the two scales and the two additional questions discussed earlier in section~\ref{sec:measures}.
The entire experiment takes \SIrange{40}{50}{\minute}.

\subsection{Sample}

\begin{filecontents*}{data/participants.csv}
participants,males,females,meanAge,minAge,maxAge,meanInteracting,meanProgramming,mean
16,11,5,33.4,23,60,4.3,3.8,1.9
\end{filecontents*}
\csvreader[head to column names]{data/participants.csv}{}{We recruited \participants\ participants (\females\ female; \males\ male) mostly from university staff and students, between the ages of \minAge\ and \maxAge\ years ($M=\meanAge$).
The participants mostly have a background in Computer Science and where asked about how familiar they are with interacting with robots, programming robots and the chosen robot platform.
A 5-point Likert scale was chosen with the value 1 for ``not familiar'' and 5 for ``very familiar''.
The self-assessed experience for interacting with robots was an average of \meanInteracting. The average familiarity with programming robots was \meanProgramming\ and the experience with the chosen robot platform was rated an average of \mean.
}
The selection of this group was on purpose, as it was assumed that this group have a more realistic view on robot's capabilities in general.
However, all participants were na\"ive with regards to the purpose of the
experiment.

The study is ethically approved by the Health, Science, Engineering \& Technology ECDA with protocol number aCOM/PGR/UH/03018.
The anonymity and confidentiality of the individual data is guaranteed.

\section{Results}
\label{sec:results}
We decided to focus on non-parametric tests to analyse the questionnaire data, as these tests are more robust for small sample sizes.
For each factor, we investigate whether there is any interaction between the order and the condition.
In case of interaction effects (i.e. order and condition) between these factors, investigating the main effects independently would be incomplete or even misleading. 

Our study can be expressed as a F1~LD~F1 Model with one within factor (condition) and one between factor (order).
A non-parametric ANOVA-type test shows that there is no interaction between the condition and the order ($p>.05$).

Participants were not explicitly asked for differences in the seen behaviour. However, we looked into the answers of the open ended questions to see if they spotted and named differences. All participants answered them, however, answers differed in detail and length.
Firstly, we checked whether participants described the behaviour and/or exploring strategy differently to the first interaction.
14 participants described the behaviour and or strategy different to the first one. Mostly, it was pointed out that one robot follows the edges more than the other or that one robot tried to leave the arena more often.
In addition, we checked whether the participants used terms like ``less'', ``compared to'' or ``this one'' for directly addressing changes to the first session after the second session. 8 out of 16 participants did so.

One participant mentioned being ``unsure, if something was fundamentally different'' but described the robots differently.
We conclude to reject hypothesis~\ref{itm:H_distinguishable} and accept the alternative hypothesis that participants were able to distinguish between the behaviours.

\csvnames{pValueReportNames0A}{factor=\factor, AMedian=\medianA, BMedian=\medianB, ABLowerConfIntWT=\CIlower, ABUpperConfIntWT=\CIupper, ABpValueWT=\p, ABEffectsizeZ=\ef}
\csvstyle{pValueReport}{
  tabular=lllll, 
  respect underscore = true,
  table head=
  \toprule 
  & \multicolumn{2}{c}{95\% confidence interval} & \multicolumn{2}{c}{}\\
  factor & \small{lower bound} & \small{upper bound} & $p$ & $r$ \\
  \midrule, 
  late after line=\\,
  late after last line=\\\bottomrule,  
  pValueReportNames0A}
  \setlength{\tabcolsep}{6pt}
\begin{table}%
    \centering
    \caption{Wilcoxon Signed Rank Test results between $REA$ and $ADA$}
    \label{tab:MIN_MAX_independent}
    \begin{filecontents*}{data/qst-rounded.csv}
factor,0Median,AMedian,BMedian,0ApValueTT,0ALowerConfIntWT,0AUpperConfIntWT,0ApValueWT,0AEffectsizeZ,0BpValueTT,0BLowerConfIntWT,0BUpperConfIntWT,0BpValueWT,0BEffectsizeZ,ABpValueTT,ABLowerConfIntWT,ABUpperConfIntWT,ABpValueWT,ABEffectsizeZ
Anthropomorphism,2.2,2.8,2.6,0.086,-0.8,0.2,0.148,0.51,0.085,-0.8,0.2,0.102,0.58,0.846,-0.3,0.4,0.916,0.037
Animacy,2.7,3.3,3.2,0.058,-0.917,0.083,0.059,0.67,0.021,-0.75,0,0.042,0.72,0.838,-0.25,0.333,0.69,0.141
Likeability,3.9,3.8,3.7,0.73,-0.6,0.7,1,0,0.516,-0.4,0.6,0.729,0.12,0.612,-0.3,0.4,0.726,0.124
Perceived Intelligence,2.7,3.2,2.8,0.245,-0.7,0.2,0.293,0.37,0.857,-0.4,0.4,0.972,0.01,0.248,-0.2,0.8,0.244,0.412
Perceived Safety,3.7,3.7,3.5,0.708,-0.667,0.5,0.677,0.15,0.864,-0.5,0.333,0.759,0.11,0.894,-0.667,0.667,0.444,0.271
Warmth,2.1,2.5,2.5,0.679,-1,0.583,0.777,0.1,0.328,-1.083,0.333,0.326,0.35,0.309,-0.667,0.167,0.366,0.32
Competence,3.3,3.7,3.4,0.569,-0.917,0.5,0.272,0.39,0.271,-1.083,0.333,0.379,0.31,0.716,-0.583,0.5,0.798,0.09
Discomfort,1.5,1.8,1.9,0.024,-1.25,-0.083,0.031,0.76,0.007,-1.75,-0.25,0.008,0.93,0.073,-0.833,0.083,0.141,0.52
\end{filecontents*}
\csvreader[pValueReport]{data/qst-rounded.csv}{}%
    {\small{\factor} & \CIlower & \CIupper &\p & \ef }%
\end{table}%
\noindent The effect size $r$ is a robust measure for small sample sizes present in this study.
The underlying effect is either small ($r=.10$), medium ($r=.30$) or large ($r=.50$)~\cite{Cohen-92}. Table~\ref{tab:MIN_MAX_independent} shows the $p$-values and effect sizes $r$ for all factors computed with the Wilcoxon Signed Rank Test.
It can be seen that there is no statistical significance for any factor. 

For the factor Animacy there is only a small effect and thus we accept the hpothesis~\ref{itm:H_change_animacy}, i.e., there is no change for the factor Animacy between both conditions.
The factor competence doesn't show any effect between the conditions and we accept the null hypothesis~\ref{itm:H_change_comp} too.

The Godspeed factor Perceived Intelligence, on the other hand, has a medium effect. We thus can reject the hypothesis~\ref{itm:H_change_PInt}. From the confidence interval we infer that this effect is in favour for the $REA$ condition. 

Interestingly, there is a medium effect for Warmth ($r>0.32$) and a large effect for Discomfort ($r>0.52$). Both factors are from the RoSA scale and both are in favour for $ADA$.

\section{Discussion}
\label{sec:conclusion_and_contribution}

It is promising that participants could see differences in the behaviour of the reactive and the adaptive behaviour. One concern with choosing a very simplistic platform is that the magnitude of behaviour differences can be very low. However, to our surprise, the perception of the robot's competence and its animal-likeness was not (significantly) increased for the adaptive robot ($ADA$) compared to the reactive one ($REA$).
We think this is due to the experimental design. Having the robot adaptive enough to leave the pit quickly was not as exciting for the participants as for us, and the quick adaptation needed to achieve this made the robot too unpredictable.
Some participants even preferred the reactive behaviour because of its predictable and stable pattern. 
Only two participants mentioned in the open ended questions that the reactive robot's trajectory was mostly circling. 
We assume that a lower update rate for the adaptive robot will make a difference here and we will address this in another study.
What also plays in favour for the reactive robot was the fact that it only rarely approaches the edge not enclosed by a wall, i.e., where the robot could fall off the table. 
The circling pattern makes it seem to be more alert for some participants, which in turn may have influenced their rating.
Initially, we thought that people would feel that the robot tries to approach them rather than trying to fall off the table. However, only one participant mentioned ``it may have sought attention''. We designed the interaction at the edge so the participants could experience the adaptation of the robot to interaction. 
However, the steady, stable behaviour of the reactive robot was preferred by participants, as it seemed easier to keep the robot from rolling over. 
This is is because the reactive robot approaches the edge less, and the robot's reactions were more predictive. 
These may all be reasons for the higher score of Perceived Intelligence for the reactive robot. We also have to rethink the initial hypothesis altogether. At the end of the day, the adaptive robot is exploring its sensor space. There is no goal other than exploring, making it more likely that the adaptive robot is indeed perceived as less competent and intelligent.

We assume a different experiment introduction could have also made a difference already. Rather than saying ``also, your task is to prevent the robot from falling off the table'', we could have said that ``we did not enclose the edge by a wall, so you can better interact with the robot when it is seeking attention, but please take care it is not rolling over the edge''.
However, this would have induced a bias towards the robot's capabilities. We think we rather need to redesign the environment for future experiments.

Although the results are not overwhelmingly convincing for Predictive Information at first glance, they have to be interpreted in the context of their novelty.
Additionally, the unexpected medium effect for Warmth is promising. The factor is created from the items Happy, Feeling, Social, Organic, Compassionate and Emotional and it carries more weight in interpersonal interaction than Competence~\cite{CarpinellaWymanEtAl-17}.

\section{Conclusion and Future Work}
In this work we used a minimal robot platform with only proprioceptive sensors updating the PI model. This makes it hard for the model to infer whether perturbations are induced by, e.g., a participant's hand or an obstacle like a wall. In other words, adaptivity to the environment and participants is limited by design. In addition, 4 participants mentioned in their open ended answers that they did not see the robot having any ``memory'' of previous obstacles or explored areas.
We want to address this with feeding other sensors into the model in future work. One option is an odometry encoder providing the robot with information about its position in the environment. 
Another option is using a previously developed proximity sensor for mobile robots for sensing interacting humans~\cite{ScheunemannDautenhahnEtAl-16}.
Both these sensors will be investigated to see whether they enhance the behaviour and make it more adaptive and interesting for HRI experiments.

One contribution of this work is the presented baseline behaviour for PI. The experiment shows that the reactive robot, based on a pre-trial adaptation of its networks configurations with PI, is a good enough candidate. Future experiments can be conducted with a baseline behaviour, which is pre-adapted for the very same sensors and the very same environment as the comparing adaptive PI behaviour.

It can be said that the capacity of the robot to leave the pit was not a driving factor for participants when judging the robot's competence positively. Again, the update rate allowing for that skill may have made the adaptive robot appear unnecessarily random. Further studies are needed for finding a good update rate for the robot.

The medium effect for Warmth for the adaptive robot is very promising. In a next step, we want to redesign the experimental setting, as the enforced interaction supposingly has a negative effect. Rather then enforcing the interaction with the robot, we want to implement a more game-like scenario which allows for the participants to interact when they feel the need or the joy to do so, rather than having to interact without them knowing when it is ``intended'' by the robot.
That way we can investigate if the effect for Warmth is indeed caused by the interaction with the intrinsically motivated robot.

\subsubsection*{Acknowledgements.}
CS is funded by the EU Horizon 2020 programme under the Marie Sklodowska-Curie grant 705643.

%
%

\end{document}